\documentclass[11pt, a4paper]{article}

\title{Profit Puzzles or: Public Firm Profits Have Fallen}
\author{\begin{tabular}{c} Carter Davis \\ Indiana University \end{tabular}  \begin{tabular}{c} Alexandre Sollaci \\ IMF \end{tabular} \begin{tabular}{c} James Traina \\ University of Chicago \end{tabular}}
\date{\today}
\usepackage[table]{xcolor}
\definecolor{darkblue}{rgb}{0.0,0.0,0.66}  
\usepackage[hyperfootnotes=false,bookmarksopen]{hyperref}
\hypersetup{
	pdffitwindow=false,
	pdfstartview={XYZ null null 1.00},
	pdfnewwindow=true,
	colorlinks=true,
	linkcolor=black,
	citecolor=darkblue,
	urlcolor=darkblue  
	} 

\usepackage{amsthm}
\usepackage{amsmath}
\usepackage{mathtools}
\usepackage{amssymb} % packages that allow mathematical formatting
\usepackage{graphicx} % package that allows you to include graphics
\usepackage{setspace} % package that allows you to change spacing
\usepackage{fullpage} % package that specifies normal margins
\usepackage{float} % moving images around
\usepackage[colorinlistoftodos]{todonotes} % comments
\usepackage{booktabs} % for nice looking tables
\usepackage{titlesec} % formatting of sections
\usepackage{ragged2e}
\usepackage[margin=1in]{geometry}
\usepackage{natbib}
\usepackage{tikz} % for timeline
\usetikzlibrary{decorations.pathreplacing}
\usepackage[bottom]{footmisc} % footnote at bottom
\usepackage{longtable}
\usepackage{bbm}
\usepackage{mathrsfs}

\newcounter{subcaption}[figure]
\newcommand{\subcaption}[1]{\refstepcounter{subcaption}\centering \textit{Panel \Alph{subcaption}: #1}}
\newcounter{subsubcaption}[subcaption]

\usepackage{dsfont} % bold 1

\renewcommand{\thesection}{\Roman{section}}

\titleformat{\section}{\Large\bfseries\centering}{\thesection}{1em}{}
\usepackage{array}
\usepackage{multirow}

\newcommand\MyBox[2]{
	\fbox{\lower0.75cm
		\vbox to 1.7cm{\vfil
			\hbox to 1.7cm{\hfil\parbox{1.4cm}{#1\\#2}\hfil}
			\vfil}%
	}%
}

\begin{document}
	\maketitle
	\thispagestyle{empty}
	\begin{center}
	PRELIMINARY AND INCOMPLETE
	\end{center}
	\doublespacing
	\begin{abstract}
	We show that public firm profit rates fell by half since 1980. Inferred as the residual from the rise of US corporate profit rates in aggregate data, private firm profit rates doubled since 1980. Public firm financial returns matched their fall in profit rates, while public firm representativeness increased from 30\% to 60\% of the US capital stock. These results imply that time-varying selection biases in extrapolating public firms to the aggregate economy can be severe. %We highlight a key implication for secular trends research in macroeconomics and finance: financial returns increasingly underestimate the aggregate cost of capital.
	\end{abstract}
	
	\clearpage
	\setcounter{page}{1}

\section{Introduction}

We compare measures of the profitability of firms in the Compustat database (which covers all non-financial publicly traded businesses in the US) with similar measures for the aggregate economy, taken from national accounts. We show that the profit-to-capital ratio for public firms has peaked in the 1980's, but has been steadily declining since then; in contrast, the corresponding aggregate measure has followed a relatively flat trend in the past 40 years -- indicating that private firms have thus been on an upward trend.

We also show that the representativess of public firms has varied widely over time: firms in Compustat owned about 30\% of all capital in the US in the 1970's, but this ratio as approximately doubled by 2020. Each of our findings is robust to different measures of the profit rate and returns on capital, as well as different weights used to derive data for private firms. We also show that our results are not driven by increased cash holdings by public firms \citep{Opleretal1999} nor by changes in the amount of goodwill.

Both issues above have important implications for research in macroeconomic and finance alike. Firm level data allows us to explore information that is simply not available at the aggregate level. As such, it is common practice for researchers to extrapolate results from data on public firms to the aggregate economy (e.g., measures of market power, discount rates, and others). We caution against such practices, as extrapolating data from public companies to the universe of firms can introduce severe bias in aggregate measures, even for something as simple as profit rates.

Our results contribute to a couple of different strands of the economic literature. First, our findings are intimately related to the debate on the rise of market power and aggregate profits \citep[e.g.,]{DeLoeckeretal2020, Barkai2020}, both of which imply that the aggregate profit ratio is increasing. We show that these trends, if true, must be driven by private firms in the US \citep[see also][]{Traina2018}. Second, we add to the evidence that public and private firms behave in distinct ways \citep{Davisetal2006, Alietal2008}, with important implications for aggregate trends.

This paper is organized as follows. Section \ref{sec:methods} lays out our methods, describing the data we use and all of the relevant calculations. Section \ref{sec:results} discusses our main results and robustness checks. Section \ref{sec:conclusion} concludes.

\section{Methods}
\label{sec:methods}

This section is laid out as follows. We first describe the data and then how the relevant economic quantities are measured in the data. Finally, we discuss the relevant filters we use for the data to obtain the aggregate results. 

\subsection{Data}

There are two main sources of data used this this paper: (1) the Bureau of Economic Analysis (BEA) data, and (2) Compustat merged with Center for Research in Security Prices (CRSP) stock data. 

We follow \cite{Barkai2020} in the definitions of variables taken from the BEA National Income and Productivity Accounts (NIPA) data tables. Annual aggregate data that contains taxes, gross value added, and labor compensation comes from NIPA Table 1.14. We use nonfinancial corporate business gross value added (line 17) as our measure of value added. Compensation of employees (line 20) measures labor costs, aggregates salaries, wages, insurance and pension contributions, most stock options, and other supplements to wage and salary income. We use taxes on production and imports less subsidies (line 23) as our measure of taxes. Finally, capital is taken from the BEA Fixed Asset Table 1.1, and private nonresidential assets (Line 4) aggregates measures of equipment, structures, and intellectual property products. 

In addition, we use data from Compustat (through WRDS) merged with CRSP to obtain stock price data. We use the standard link types: links verified by CRSP (LC links) and links based on asset Committee on Uniform Securities Identification Procedures (CUSIP) numbers (LU links). Primary issue markers are used for the links between Compustat and CRSP (P and C). To match the macro data, the observation year is the calendar year of the end of fiscal year for the reporting firm. In the few instances where a firm reports multiple annual results in one calendar year, the last observation in the given calendar year is kept. Note that while this practice might introduce some measurement error to our data on a yearly basis, it does not affect secular trends, which is the main object of interest for this paper.

\subsection{Measurement}

For the most part in this paper, we will focus on what we call the profit ratio, or the profit-to-capital ratio. It is defined as
\[ \pi_t = \frac{X_t}{K_t}, \]
where $K_t$ is the value of the capital stock in period $t$ and $X_t$ is aggregate cash earnings (after taxes, but before interest and depreciation).Note that because this is a contemporaneous ratio, it is invariant to the various measures of inflation and other dynamic issues. Going forward, we drop the $t$ subscripts whenever it does not cause confusion.

We follow \cite{FamaFrench1999} in our definitions of variables from the CRSP/Compustat data. The measure of profits, $X$, is defined as the sum of income before extraordinary items (IB), income from extraordinary items (XIDO), deferred taxes (TXDI), interest expense (XINT), and depreciation and amortization. Depreciation and amortization is measured as the depreciation and amortization variable (DP) if available, and depreciation expense (XDP) variable otherwise. Capital $K$ is measured using book capital, defined as the sum of long term debt (DLTT), short term debt (DLC), and the book value of equity. The book value of equity equals the total assets variable (AT) minus total liabilities (LT) plus the deferred taxes and investment tax credit (TXDITC), if available. Cash is measured with the cash and short term investments variable (CHE). Goodwill is measured with the standard Compustat goodwill variable (GDWL). 

Lastly, the profit-to-capital ratio is also measured using the BEA data. US macroeconomic profits $X_{macro}$ are measured as gross value added minus taxes and labor costs. 

\subsection{Aggregation}

We drop utilities and financial firms from our CRSP/Compustat sample to better match the macroeconomic data, and because they follow different dynamics as other firms.\footnote{Utility firms are heavily regulated in many cases, which can constrain their ability to maximize profits. Financial firms' balance sheets can be much different from other firms, and are thus also removed so we can maintain consistency of our measurements. (Source: Thomas Winberry's problem set from 'Micro Data for Macro Models'; available at: \url{https://www.thomaswinberry.com/teaching/index.html}.)}
As shown in Table \ref{tbl:filters}, this cuts out about 42\% of our sample in dollars terms ($1 - 1.66 / 2.87$). 

To remove non-US data from our sample, we only keep observations where the following three conditions hold: (1) the foreign code variable equals the US (foreign\_code $=$ USA), (2) amounts are reported in US dollars, and (3) firms are listed on either the New York Stock Exchange or NASDAQ. This cuts out another sizable portion of the data, indicating that a large share of Compustat includes foreign operations. Finally, we also drop observations where profits $X$, capital $K$, or market value $V$ are missing. 
We highlight that there is on average almost three times as much book capital in nominal dollar terms in the unfiltered CRSP/Compustat data than in our filtered sample (see table \ref{tbl:filters}). This shows that a substantial amount of the Compustat data represents either foreign operations or US domestic finance or utilities firms. Because of this, using CRSP/Compustat data without filters applied to understand nonfinancial US corporate trends is problematic at best. 

\section{Results}
\label{sec:results}

In this section, we first highlight the evidence that public firm profits have fallen. We then show various measures of the private share of profits. Finally, we test the robustness of our findings against several alternative measures. 

\subsection{Public Firm Profits Have Fallen}

\begin{figure}[!h]
\caption{Compustat US Profits to Capital Ratio} \label{fig:compustat_profits}
\bigskip
\centering
\includegraphics[scale=1.0]{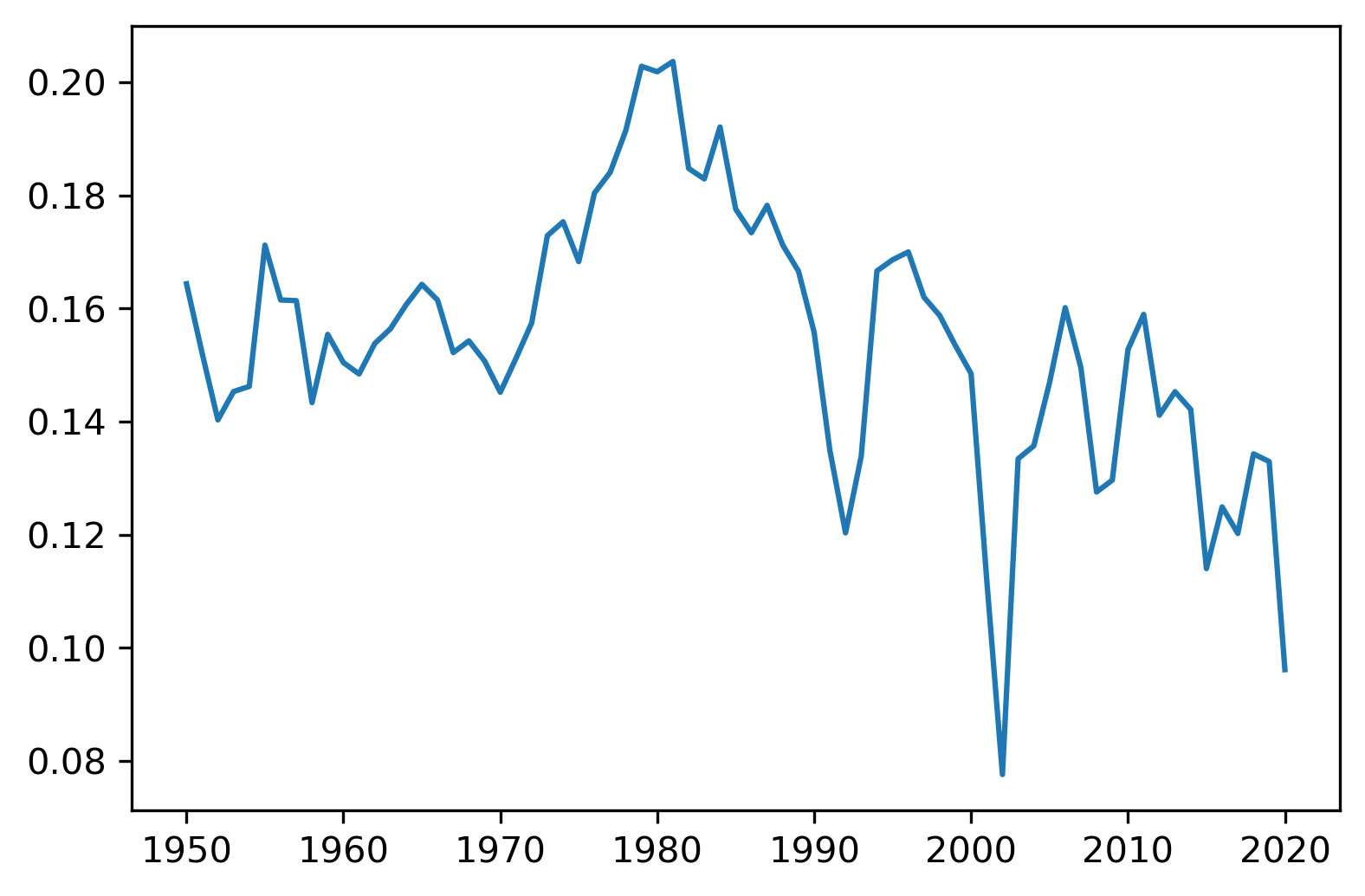}
\begin{flushleft}\singlespacing \footnotesize This plot shows US profits over capital from Compustat. Aggregate profits and capital are defined in \cite{FamaFrench1999}. Profits are defined as aggregate cash earnings after taxes but before deduction of interest and depreciation ($IB + XIDO + TXDI + XINT + DP$). Capital is defined as the sum of long term debt (DLTT), short term debt (DLC), and book equity (total assets minus total liabilities plus the deferred taxes and investment tax credit if available --- $AT - LT + TXDITC$). Finance (NAICS 520000 through 529999) and utilities (NAICS 220000 through 229999) are dropped. Only US companies are considered ($FIC = USA$, $CURCD = USD$, as well as $EXCHG = 11$ or $ECHG = 14$).
\end{flushleft}
\end{figure}

While aggregate public firm profits increased during the 1970's and early 1980's, public firm profits have trended downwards since. These results are shown in Figure \ref{fig:compustat_profits}. Profit ratios hover around 16\% in the 1950's and 1960's. In the 1970's, profit ratios spike as high as 20\%, trending downward to an average of about 12\% in the 2010's. Those findings are in line with \cite{KahleStulz2017}, who point out that the average public US corporation is less profitable than it was 40 years ago.

\begin{figure}[!h]
\caption{US Profits to Capital Ratio} \label{fig:macro_profits}
\bigskip
\centering
\includegraphics[scale=1.0]{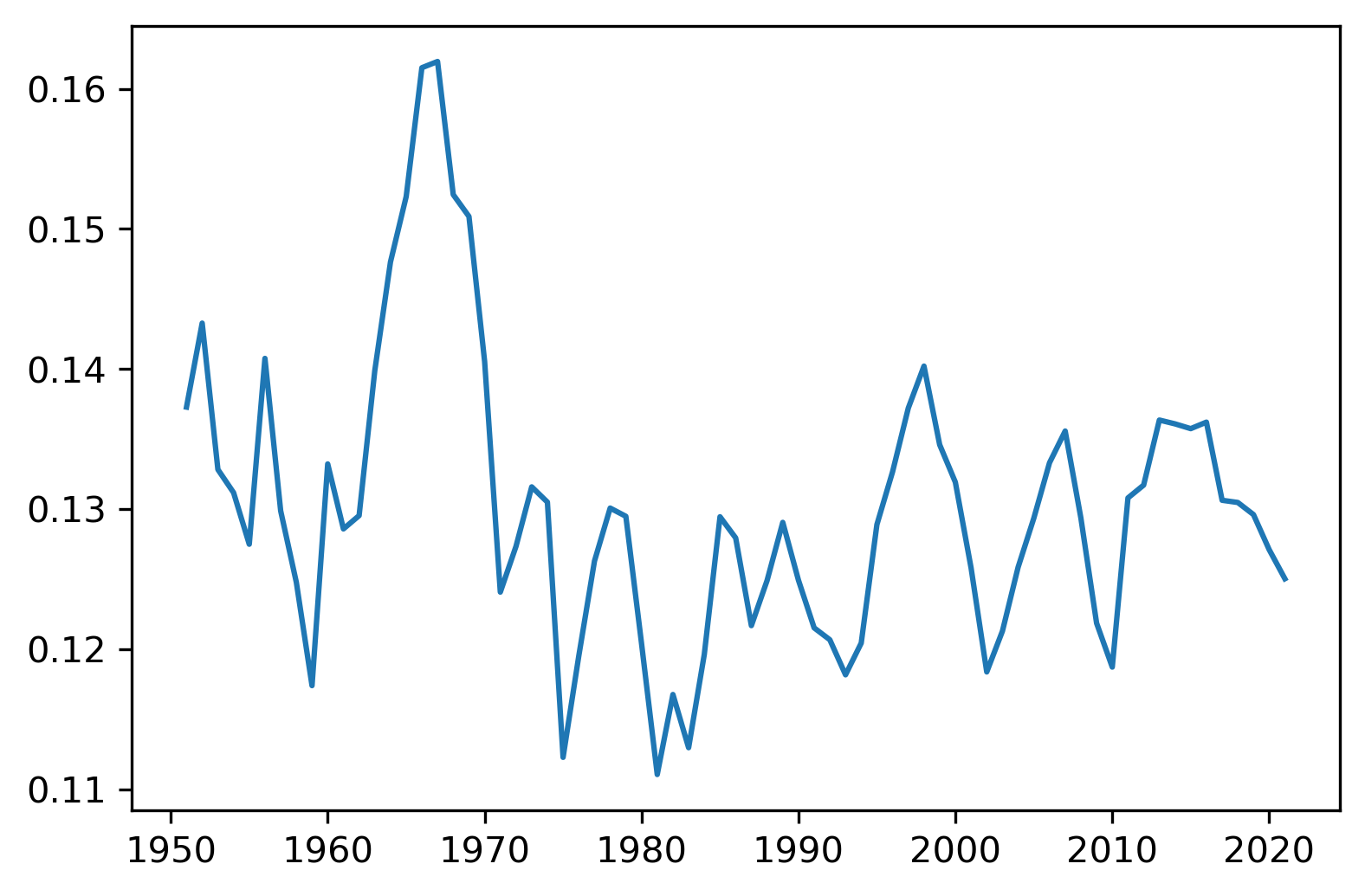}
\begin{flushleft}\singlespacing \footnotesize This plot shows US profits over capital. We follow \cite{Barkai2020} in how we compute profits and capital. Profits come from National Income and Product Accounts (NIPA) Table 1.14, and are defined as gross value added of nonfinancial corporate business (line 17), minus compensation of employees (line 20) and taxes on production and imports less subsidies (line 23). Capital comes from the BEA fixed asset table 1.1. Capital is defined as nonresidential private capital (line 4). 
\end{flushleft}
\end{figure}

In contrast, the aggregate macroeconomic profit ratio is shown in Figure \ref{fig:macro_profits}. While US aggregate firm profits spiked early in the sample, the trend after the 1970's is relatively flat. This implies that private firm profits must have risen over this period since private firm profits and public firm profits must sum to aggregate US firm profits (see section \ref{sec:private_profits}).

This finding has important implications for a whole host of facts documented about rising markups, market concentration, and profits in the US \citep[e.g.,][]{Basu2019, DeLoeckeretal2020, Grullonetal2019, Traina2018}. In particular, if markups have increased over the past decades, a similar increase in profits should have been observed as well. Our results suggest that this hasn't been the case for public firms in the US, pointing to either private or non-US firms as the drivers of this trend.\footnote{\cite{Diezetal2021} document an increase in markups for a large sample of private and listed firms from both advanced economies and emerging markets.}

On the rise of profits, \cite{Barkai2020} argues that both the labor and capital share of value added have decreased since the 1980's, implying the the profit share must have increased. Defining the capital and profit shares as $rK/Y$ and $X/Y$ (where $Y$ is gross value added and $r$ is the interest rate), it is clear that $X/(rK)$ must have increased on aggregate. In contrast, our findings suggest that aggregate $X/K$ has followed a relatively flat trend since the 1970's, meaning that the trend found by \cite{Barkai2020} is mostly driven by his measure of the interest rate -- an argument originally made by \cite{KarabarbounisNeiman2019}. 

In a related point, we show in appendix \ref{app:irr} that our measure of the profit ratio is closely related to the 1-period internal rate of return on capital. For some applications, this might be a better measure of the interest rate on aggregate capital than, say, the yield on government bonds.\footnote{As \cite{Mulligan2002} argues, "the" interest rate in aggregate is not observable: it is not the yield on a Treasury Bill or Bond, but the expected return on a unit of representative capital.} However, unlike the yield on bonds or related measures, which have drastically fallen since the 1980's, we find that the aggregate profit ratio (and thus the IRR on capital) has -- as mentioned above -- had a flat trend over the same period. 

Finally, we add that our results are not driven by the specific measure of capital we use. \cite{CrouzetEberly2021} stress the importance of appropriately measuring intangible capital -- which could be underestimated if we measure capital by its book vale. In contrast, the market value of capital is more likely to capture the value of intangible capital \citep{Hall2001}, as well as market power and other firm characteristics that might affect profitability. To that end, Figure \ref{fig:compustat_market_profits} shows the profit to market capital ratio, $X / V$. This shows an even stronger decline than the profit to capital ratio. Thus, if capital is measured as market capital, profits appear to decline even more precipitously over this period. 

\subsection{Compustat Share of Aggregate Capital has Increased}

As described above, we have measures of aggregate US capital (which we denote as $K_{\text{macro}}$) from the BEA data and measures of public US firm capital from Compustat data (denoted as $K_{\text{Compustat}}$). We can take this ratio, $w = K_{\text{Compustat}} / K_{\text{macro}}$, to determine the fraction of capital in the Compustat sample to the US aggregate capital. 

\begin{figure}[!h]
\caption{US Compustat Capital to Aggregate Capital} 
\label{fig:capital_ratio}
\bigskip
\centering
\includegraphics[scale=1.0]{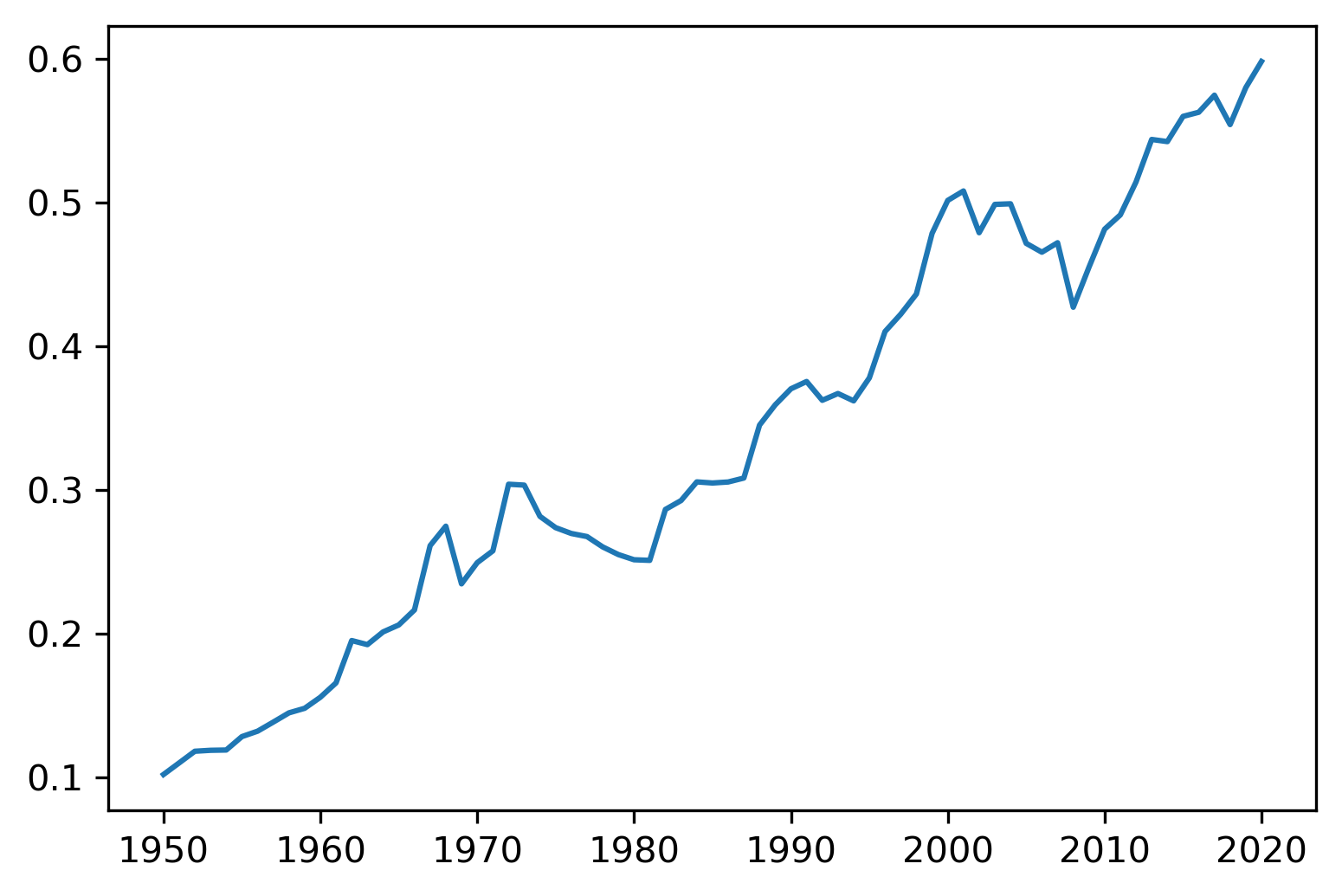}
\begin{flushleft}\singlespacing \footnotesize This plots shows the ratio of Compustat capital to US aggregate capital. Descriptions of the way that capital is calculated from the two datasets is contained in the two plots above. 
\end{flushleft}
\end{figure}

The ratio of Compustat capital to US macroeconomic capital is shown in Figure \ref{fig:capital_ratio}. This ratio is initially around 0.1 when the Compustat is relatively sparse in the 1950's. In the 1980's when the Compustat data is more rich, the ratio is around 0.3. This is also when the number of listed firms in the US expands dramatically \cite{FamaFrench2004}, potentially explaining the trend reversal around that time. 

The ratio then trends upwards fairly steadily until it reaches about 0.6 in 2020. A direct implication of this observation is that, as measured by its share of aggregate capital, Compustat data have become increasingly representative over our sample period. Curiously, the last portion of the increase in the share of aggregate capital represented in Compustat happens concomitantly with a decrease in the number of listed firms in the US, suggesting that public firms have also become bigger over time \citep{Doidgeetal2017, KahleStulz2017}.

A note of caution is deserved at this point. Compustat book capital and the BEA fixed asset tables have a multitude of different accounting treatments. This is certainly true for profit measures across the two datasets. We acknowledge that the measurement error is potentially large. However, this exercise remains interesting to understand how Compustat data trends differ than macroeconomic data. Cost of capital rates and various other economic quantities are often computed in Compustat and imputed to the rest of the US aggregate data. Given the common use of Compustat data to impute these important economic quantities to the macroeconomic data, this comparison is in order.

\subsection{Implied Private Firm Profits have Increased}
\label{sec:private_profits}

Using the profit ratio series and the Compustat capital to BEA capital ratio, $w$, we can compute the implied private firm profit ratio. The private firm profit ratio, which we write as $X_{\text{residual}} / K_{\text{residual}}$, can be found by simply setting $X_{\text{residual}} = X_{\text{macro}} - X_{\text{Compustat}}$ and $K_{\text{residual}} = K_{\text{macro}} - K_{\text{Compustat}}$. Equivalently, we can solve the following equation for $X_{\text{residual}} / K_{\text{residual}}$:
\begin{equation*}
    \frac{X_{\text{macro}}}{K_{\text{macro}}} = w \frac{X_{\text{Compustat}}}{K_{\text{Compustat}}} + (1 - w) \frac{X_{\text{residual}}}{K_{\text{residual}}}
\end{equation*}

\begin{figure}[!h]
\caption{Profit to Capital Ratios} \label{fig:private_profits}
\bigskip
\centering
\includegraphics[scale=1.0]{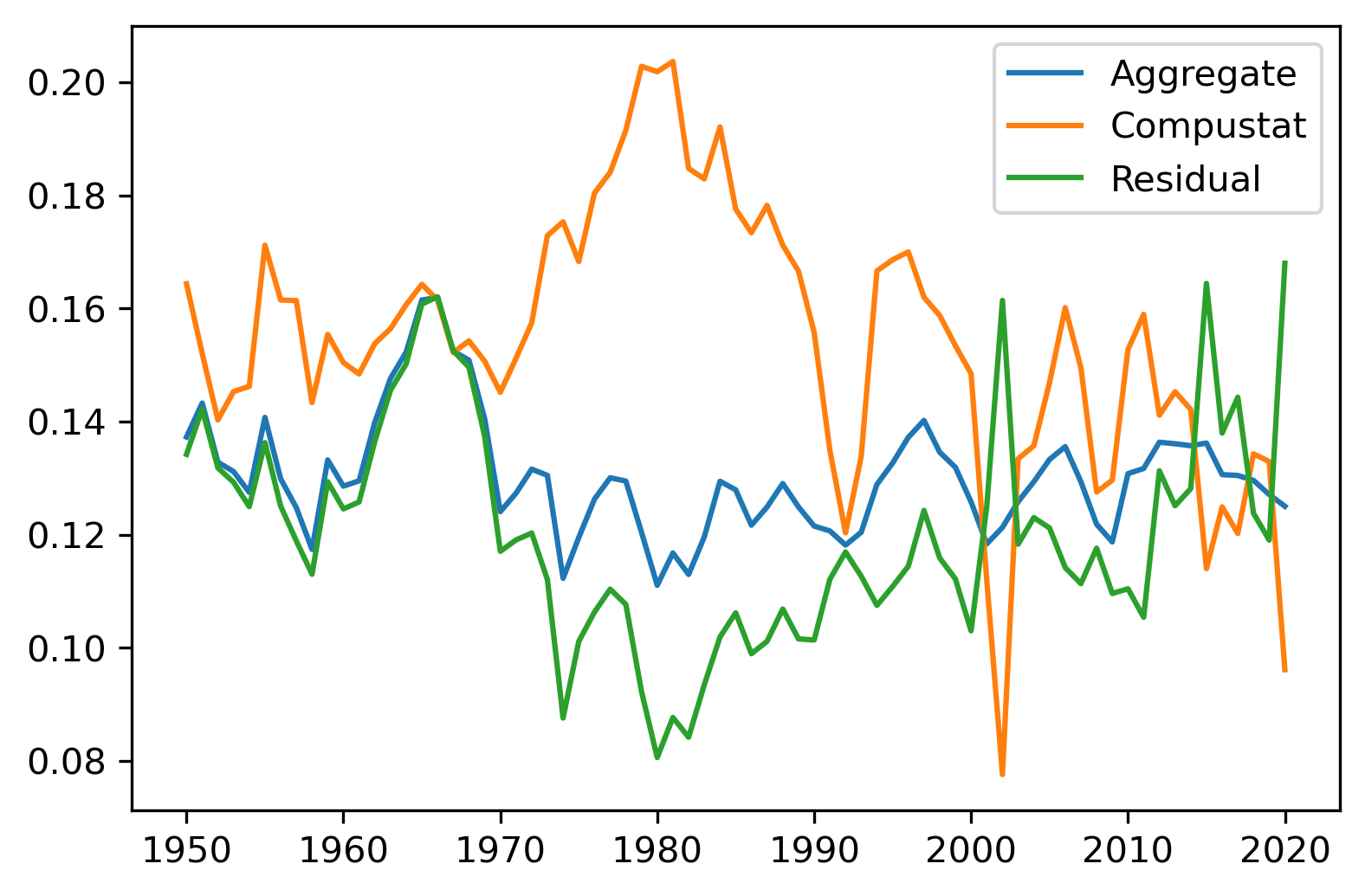}
\begin{flushleft}\singlespacing \footnotesize This plot shows the profits to capital ratio of in the aggregate US data, in the Compustat data, and the residual profits to capital ratio of firms in the US but outside of the Compustat data. Let $X$ denote profits, $K$ denote capital, and thus $X_{\text{residual}} / K_{\text{residual}}$ is calculated by solving
\begin{equation*}
    \frac{X_{\text{macro}}}{K_{\text{macro}}} = w \frac{X_{\text{Compustat}}}{K_{\text{Compustat}}} + (1 - w) \frac{X_{\text{residual}}}{K_{\text{residual}}}
\end{equation*}
where $w$ is the Compustat to aggregate capital ratio derived from the plot immediately above. 
\end{flushleft}
\end{figure}

Figure \ref{fig:private_profits} shows all three profit ratios in one graph. As expected, private firm profits trend upwards while public firm profits trend downwards. Aggregate, public, and private profits thus have been much closer together in the past few decades than during the pre-1990 period, appearing to have converged over the 1980's. Our observation that public and private firms' profit rates have followed different trends adds to the evidence that the two categories of firms are fundamentally different. Among others, \cite{Davisetal2006} document that the dispersion in growth rates of public firms is much larger than that of private firms, while \cite{Alietal2008} and \cite{Keil2017} show that industry concentration measures using public firms are essentially uncorrelated with the same measures using Census data.

As an alternative check, we can plug in other values for $w$ and solve for $X_{\text{residual}} / K_{\text{residual}}$ using the equation above. Figure \ref{fig:alt_private_profits} shows the results of this exercise. Panel A shows the results with $w = 0.3$, and Panel B shows the three series with $w = 0.5$. While the levels vary between the plots, the general trend of falling public firm profits, rising private firm profits, and general convergence holds in every case. 

\subsection{Robustness Checks}

We test whether our results hold under different measures of returns of capital. To that end, we follow \cite{FamaFrench1999} to define a measure of the return on the market value of firms, $r_V$, as well as a measure of the returns on capital, $r_K$, both discussed in detail in appendix \ref{app:irr}. The return on value is relatively volatile, given the volatility of market equity valuations. Thus we plot a 10-year moving average for both types of returns in Figure \ref{fig:book_market_returns}. Returns on capital shown in Panel A show precipitous declines. Market returns on Panel B trend downward from the mid-1990's to the mid 2000's, but increase afterward due to rising market valuations. Thus, while profits have declined, the capital gains portion of returns on value has lead to increasing returns. 

In Figure \ref{fig:capital_ratio_nocash} we plot $X / (K - \text{cash})$ in the Compustat data to determine if the fall in profits is driven by the rise in cash holdings documents by \citep[e.g.,][]{Opleretal1999}. This plot shows that the decline in public firm profits is not due to the rise in cash holdings. 

In figure \ref{fig:profit_goodwill_ratio}, we plot profits to goodwill. Unfortunately, goodwill is not available until 1988. For the first few decades, goodwill is relatively small compared to profits. While a substantial amount of goodwill has accumulated in the mid-2000's and 2010's, there simply is not enough goodwill coverage during our sample period to determine how important goodwill is to the decline in profits during our sample period.

\section{Conclusion}
\label{sec:conclusion}

In this paper, we show that public and private firms have followed different trends in the US, especially since the 1980's. In particular, we compare measures aggregate profitability from national accounts with similar measures from data on public firms. We find evidence that public firms' profit ratio has declined, while the aggregate profit ratio has remained relatively constant over the past decades -- indicating that private firms' profit ratios have increased. In addition, we also find that the share of capital owned by public firms has steadily increased over time, implying significant changes in how representative those firms are of the overall economy. 

Our findings have relevant implications for macroeconomic and financial research, where it is common practice to extrapolate measures computed from data on public firms to aggregate variables. In particular, we argue that such exercises could lead to severe biases, as variables computed from public firm data -- even as basic as profitability -- might not adequately reflect aggregate trends.

\clearpage
\bibliography{references.bib}

\begin{thebibliography}{18}
\expandafter\ifx\csname natexlab\endcsname\relax\def\natexlab#1{#1}\fi

\bibitem[Ali et~al.(2008)Ali, Klasa, and Yeung]{Alietal2008}
Ali, Ashiq, Sandy Klasa, and Eric Yeung, 2008, {The Limitations of Industry
  Concentration Measures Constructed with Compustat Data: Implications for
  Finance Research}, {\em The Review of Financial Studies\/} 22, 3839--3871.

\bibitem[Barkai(2020)]{Barkai2020}
Barkai, Simcha, 2020, Declining labor and capital shares, {\em The Journal of
  Finance\/} forthcoming.

\bibitem[Basu(2019)]{Basu2019}
Basu, Susanto, 2019, Are price-cost markups rising in the united states? a
  discussion of the evidence, {\em Journal of Economic Perspectives\/} 33,
  3--22.

\bibitem[Crouzet and Eberly(2021)]{CrouzetEberly2021}
Crouzet, Nicolas, and Janice~C. Eberly, 2021, {Intangibles, Markups, and the
  Measurement of Productivity Growth}, NBER Working Papers 29109, National
  Bureau of Economic Research, Inc.

\bibitem[Davis et~al.(2006)Davis, Haltiwanger, Jarmin, Miranda, Foote, and
  Nagypál]{Davisetal2006}
Davis, Steven~J., John Haltiwanger, Ron Jarmin, Javier Miranda, Christopher
  Foote, and Éva Nagypál, 2006, Volatility and dispersion in business growth
  rates: Publicly traded versus privately held firms, {\em NBER Macroeconomics
  Annual\/} 21, 107--179.

\bibitem[De~Loecker et~al.(2020)De~Loecker, Eeckhout, and
  Unger]{DeLoeckeretal2020}
De~Loecker, Jan, Jan Eeckhout, and Gabriel Unger, 2020, {The Rise of Market
  Power and the Macroeconomic Implications}, {\em The Quarterly Journal of
  Economics\/} 135, 561--644.

\bibitem[Doidge et~al.(2017)Doidge, Karolyi, and Stulz]{Doidgeetal2017}
Doidge, Craig, G.~Andrew Karolyi, and René Stulz, 2017, The u.s. listing gap,
  {\em Journal of Financial Economics\/} 123, 464--487.

\bibitem[Díez et~al.(2021)Díez, Fan, and Villegas-Sánchez]{Diezetal2021}
Díez, Federico~J., Jiayue Fan, and Carolina Villegas-Sánchez, 2021, Global
  declining competition?, {\em Journal of International Economics\/} 132,
  103492.

\bibitem[Fama and French(1999)]{FamaFrench1999}
Fama, Eugene~F., and Kenneth~R. French, 1999, The corporate cost of capital and
  the return on corporate investment, {\em The Journal of Finance\/} 54,
  1939--1967.

\bibitem[Fama and French(2004)]{FamaFrench2004}
Fama, Eugene~F, and Kenneth~R French, 2004, New lists: Fundamentals and
  survival rates, {\em Journal of Financial Economics\/} 73, 229--269.

\bibitem[Grullon et~al.(2019)Grullon, Larkin, and Michaely]{Grullonetal2019}
Grullon, Gustavo, Yelena Larkin, and Roni Michaely, 2019, {Are US Industries
  Becoming More Concentrated?}, {\em Review of Finance\/} 23, 697--743.

\bibitem[Hall(2001)]{Hall2001}
Hall, Robert~E., 2001, The stock market and capital accumulation, {\em American
  Economic Review\/} 91, 1185--1202.

\bibitem[Kahle and Stulz(2017)]{KahleStulz2017}
Kahle, Kathleen~M., and René~M. Stulz, 2017, Is the us public corporation in
  trouble?, {\em Journal of Economic Perspectives\/} 31, 67--88.

\bibitem[Karabarbounis and Neiman(2019)]{KarabarbounisNeiman2019}
Karabarbounis, Loukas, and Brent Neiman, 2019, {Accounting for Factorless
  Income}, {\em NBER Macroeconomics Annual\/} 33, 167--228.

\bibitem[Keil(2017)]{Keil2017}
Keil, Jan, 2017, The trouble with approximating industry concentration from
  compustat, {\em Journal of Corporate Finance\/} 45, 467--479.

\bibitem[Mulligan(2002)]{Mulligan2002}
Mulligan, Casey~B., 2002, {Capital, Interest, and Aggregate Intertemporal
  Substitution}, NBER Working Papers 9373, National Bureau of Economic
  Research, Inc.

\bibitem[Opler et~al.(1999)Opler, Pinkowitz, Stulz, and
  Williamson]{Opleretal1999}
Opler, Tim, Lee Pinkowitz, René Stulz, and Rohan Williamson, 1999, The
  determinants and implications of corporate cash holdings, {\em Journal of
  Financial Economics\/} 52, 3--46.

\bibitem[Traina(2018)]{Traina2018}
Traina, James, 2018, {Is Aggregate Market Power Increasing? Production Trends
  Using Financial Statements}, {\em Stigler Center New Working Paper Series\/}
  17.

\end{thebibliography}

\clearpage
\appendix 

\section{Figures}
\renewcommand{\thefigure}{\thesection.\arabic{figure}}
\setcounter{figure}{0}

\begin{figure}[!h]
\caption{Compustat US Profits to Market Capital Ratio} \label{fig:compustat_market_profits}
\bigskip
\centering
\includegraphics[scale=1.0]{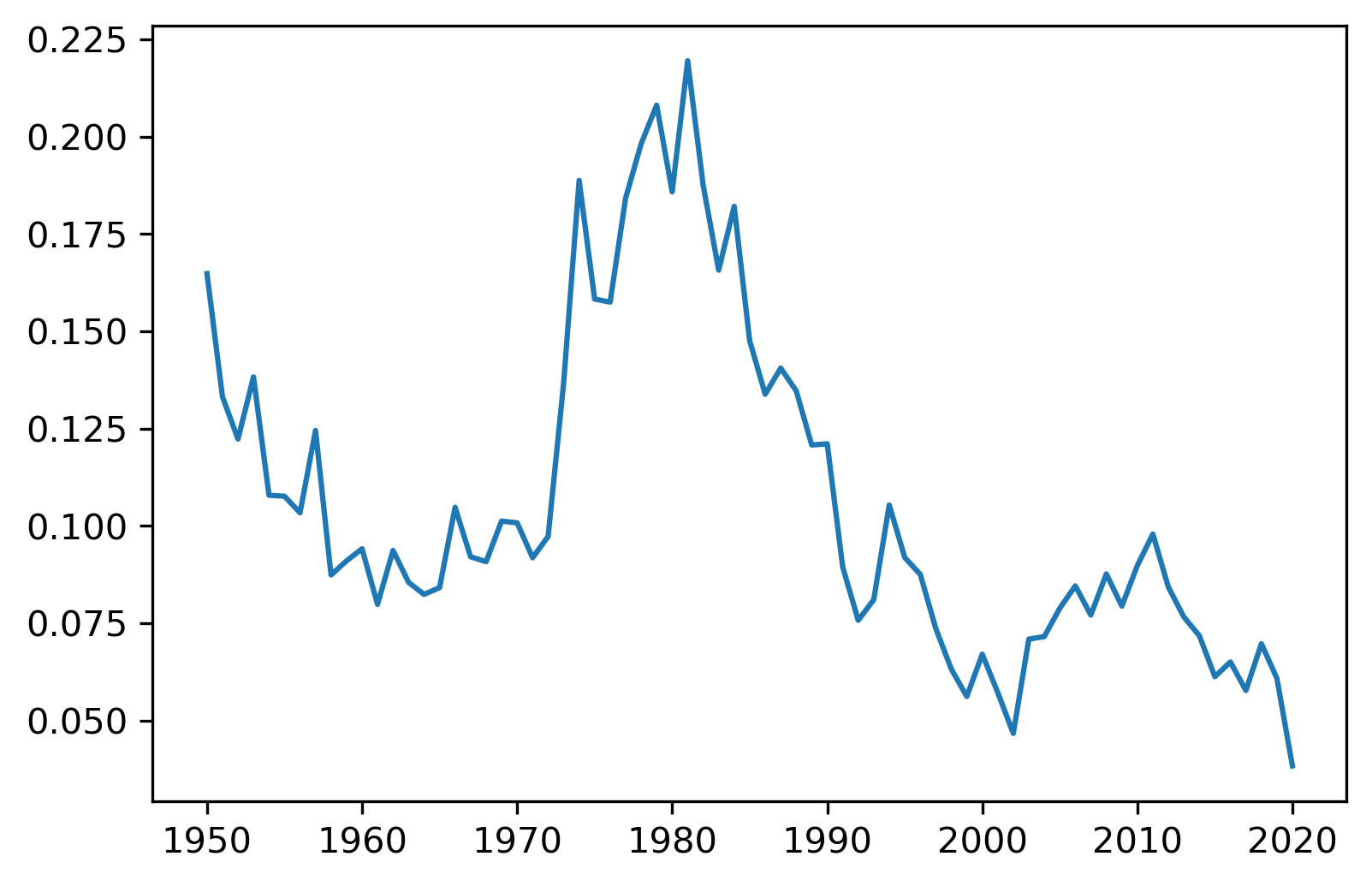}
\begin{flushleft}\singlespacing \footnotesize This plot shows US profits over market capital from Compustat. Aggregate profits and capital are defined in \cite{FamaFrench1999}. Profits are defined as aggregate cash earnings after taxes but before deduction of interest and depreciation ($IB + XIDO + TXDI + XINT + DP$). Market capital is defined as the sum of long term debt (DLTT), short term debt (DLC), and market equity. 
\end{flushleft}
\end{figure}

\begin{figure}[!h]
\caption{Alternative Profit to Capital Ratios} \label{fig:alt_private_profits}
\bigskip
\centering
\begin{minipage}{.5\textwidth}
\vspace{4mm}
   \begin{center}
   \subcaption{Profit to Capital Ratio with $w = 0.3$}
	\includegraphics[scale=0.57]{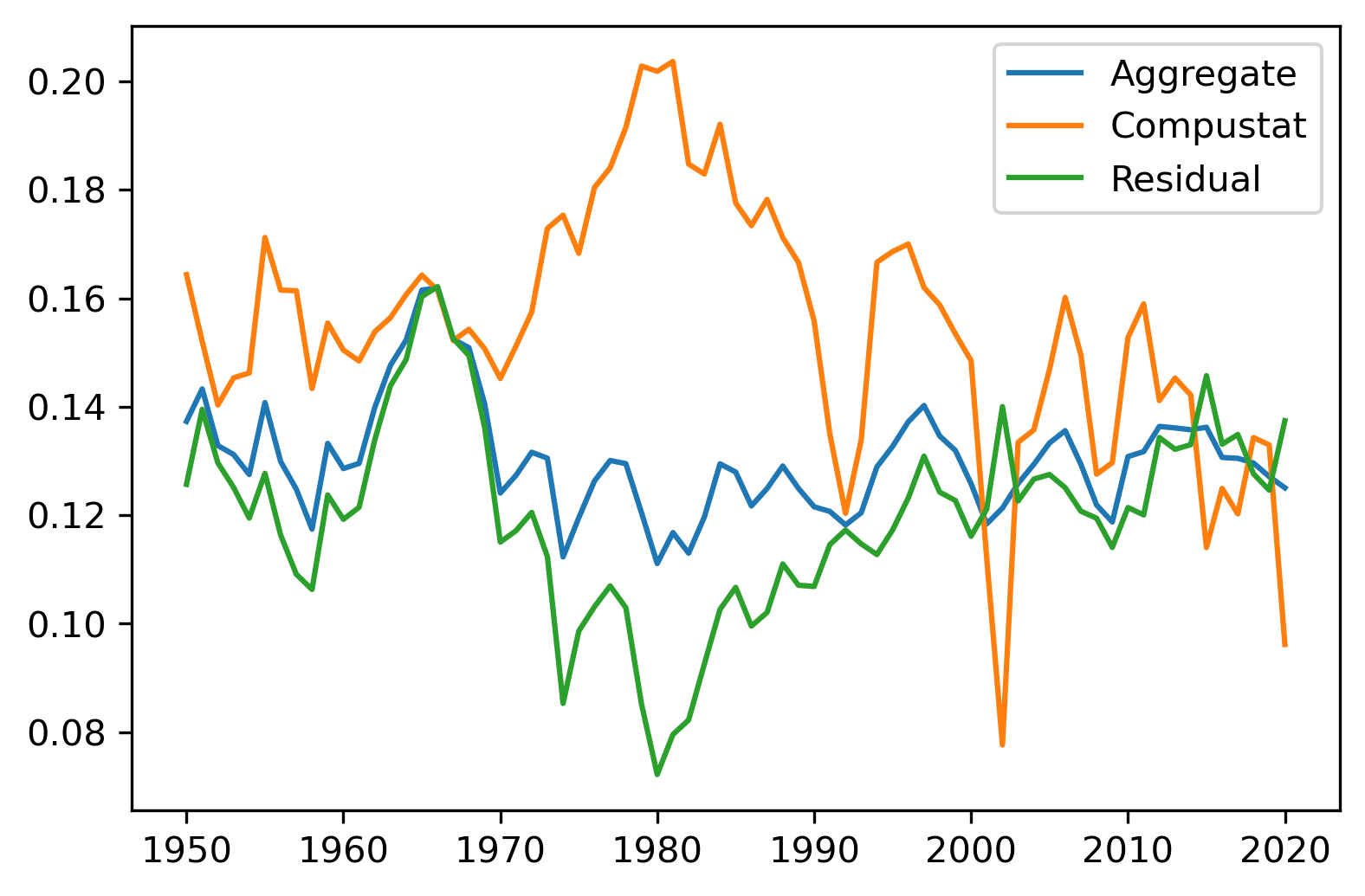}
\end{center}
\end{minipage}%
\bigskip\begin{minipage}{.5\textwidth}
\vspace{4mm}
  \begin{center}
  \subcaption{Profit to Capital Ratio with $w = 0.5$}
	\includegraphics[scale=0.57]{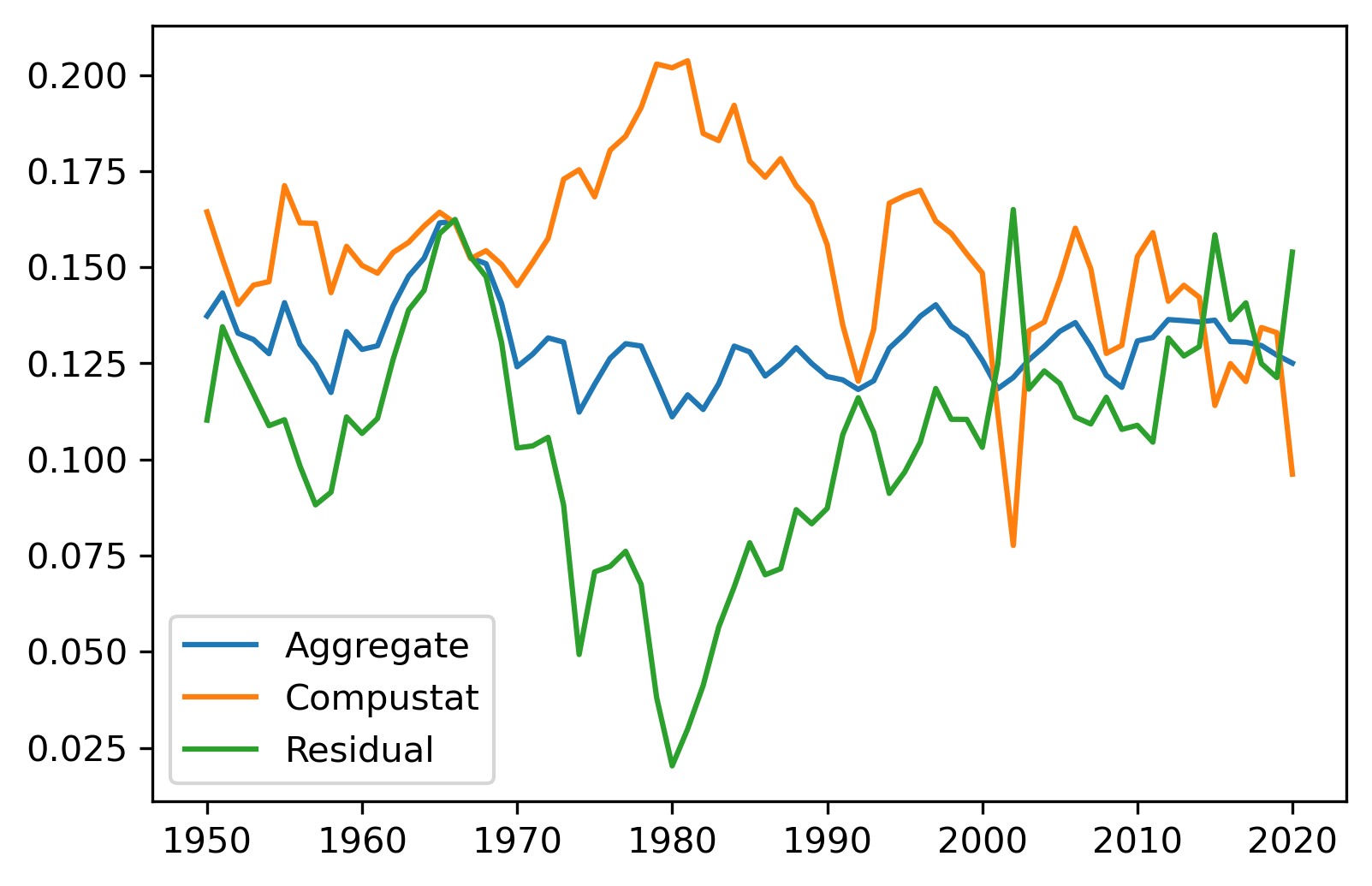}
\end{center}
\end{minipage}
\begin{flushleft}\singlespacing \footnotesize This figure plots the same profit to capital ratios from the plot above, except with $w = 0.3$ in Panel A and $w = 0.5$ in Panel B. 
\end{flushleft}
\end{figure}

\begin{figure}[!h]
\caption{Measures of Book and Market Returns} \label{fig:book_market_returns}
\bigskip
\centering
\begin{minipage}{.5\textwidth}
\vspace{4mm}
   \begin{center}
   \subcaption{Book Return}
	\includegraphics[scale=0.57]{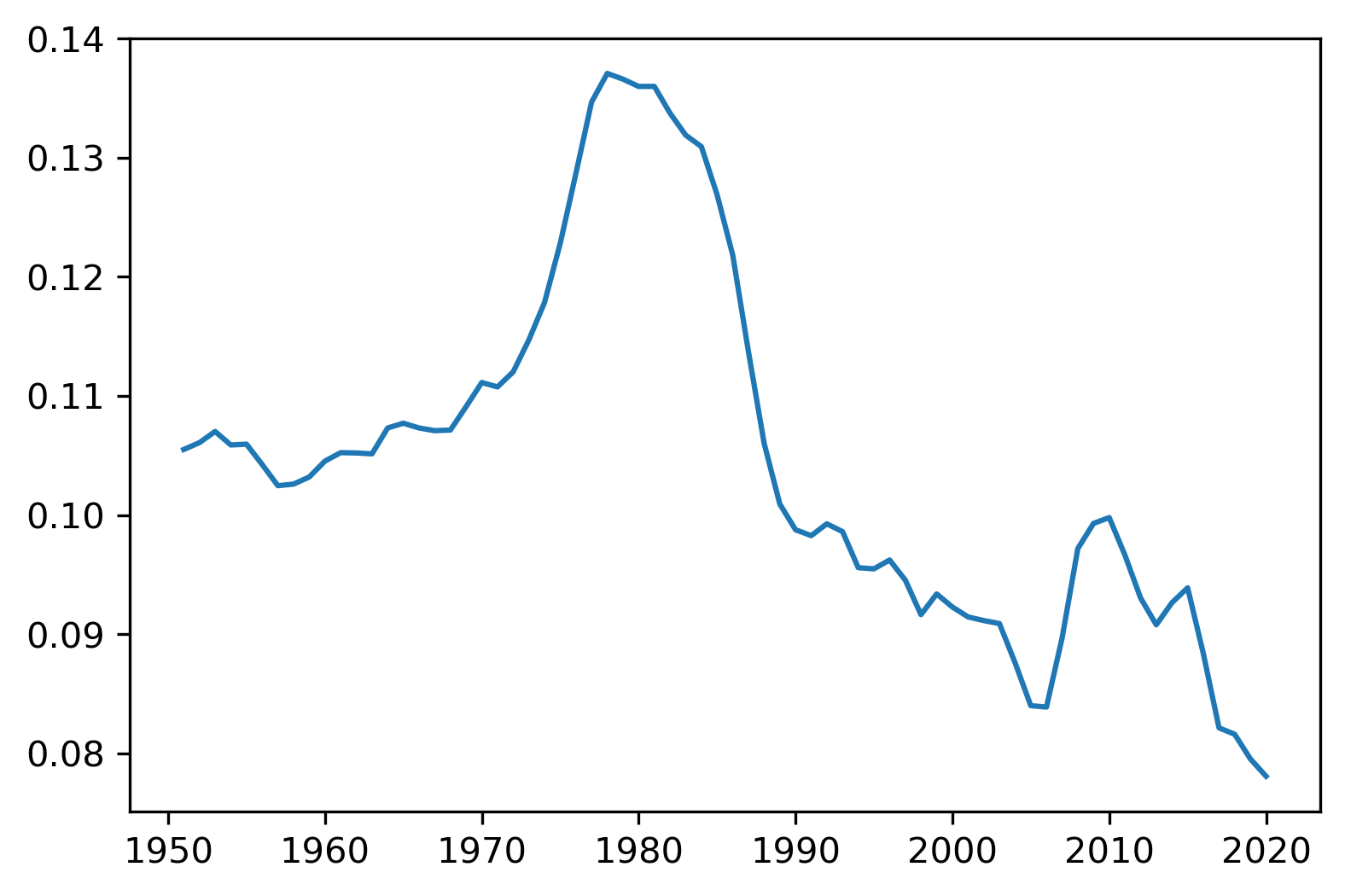}
\end{center}
\end{minipage}%
\bigskip\begin{minipage}{.5\textwidth}
\vspace{4mm}
  \begin{center}
  \subcaption{Market Return}
	\includegraphics[scale=0.57]{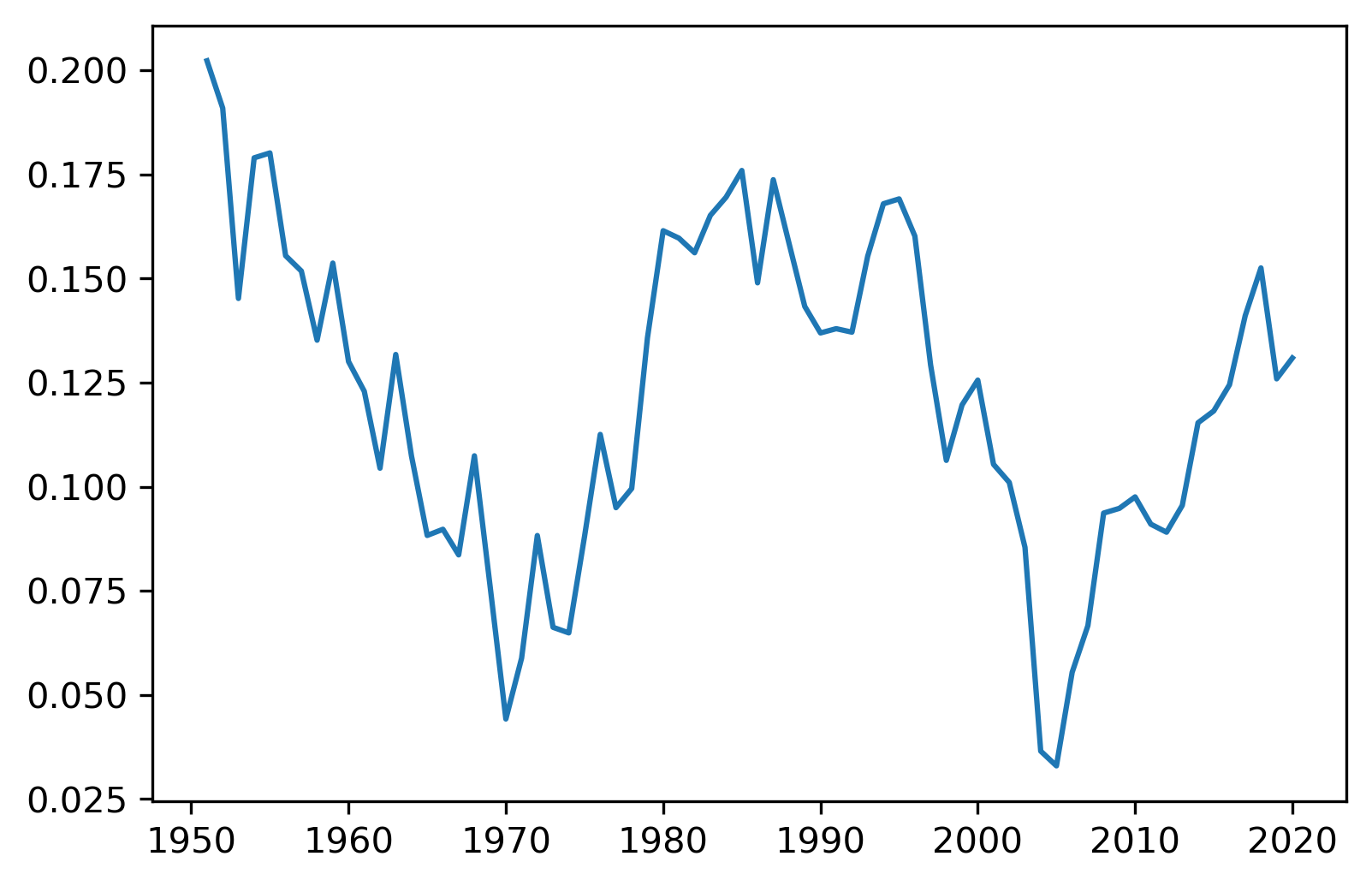}
\end{center}
\end{minipage}
\begin{flushleft}\singlespacing \footnotesize This figure plots 10-year moving averages of aggregate book returns and market returns using Compustat data, following \cite{FamaFrench1999}. Book returns ($RB_t$) are defined as
\begin{equation*}
    RB_t = \frac{X_t - I_t + K_t}{K_{t-1}} - 1 = \frac{X_t}{K_{t-1}}
\end{equation*}
where $X_t$ are profits, $I_t$ is investment, and $K_t$ is book capital. Profits are defined as aggregate cash earnings after taxes but before deduction of interest and depreciation ($IB + XIDO + TXDI + XINT + DP$). Book capital is defined as the sum of long term debt (DLTT), short term debt (DLC), and book equity. Book equity is calculated as total assets (AT) minus total liabilities (LT) plus the deferred taxes investment tax credit (TXDITC). Investment is the change in book capital. Market returns ($RV_t$) are defined as
\begin{equation*}
    RV_t = \frac{X_t - I_t + V_t}{V_{t-1}} - 1 
\end{equation*}
where $V_t$ is market capital. Market capital is defined as the sum of long term debt (DLTT), short term debt (DLC), and market equity. 
\end{flushleft}
\end{figure}

\begin{figure}[!h]
\caption{Profit to Capital Ratio without Cash} \label{fig:capital_ratio_nocash}
\bigskip
\centering
\includegraphics[scale=1.0]{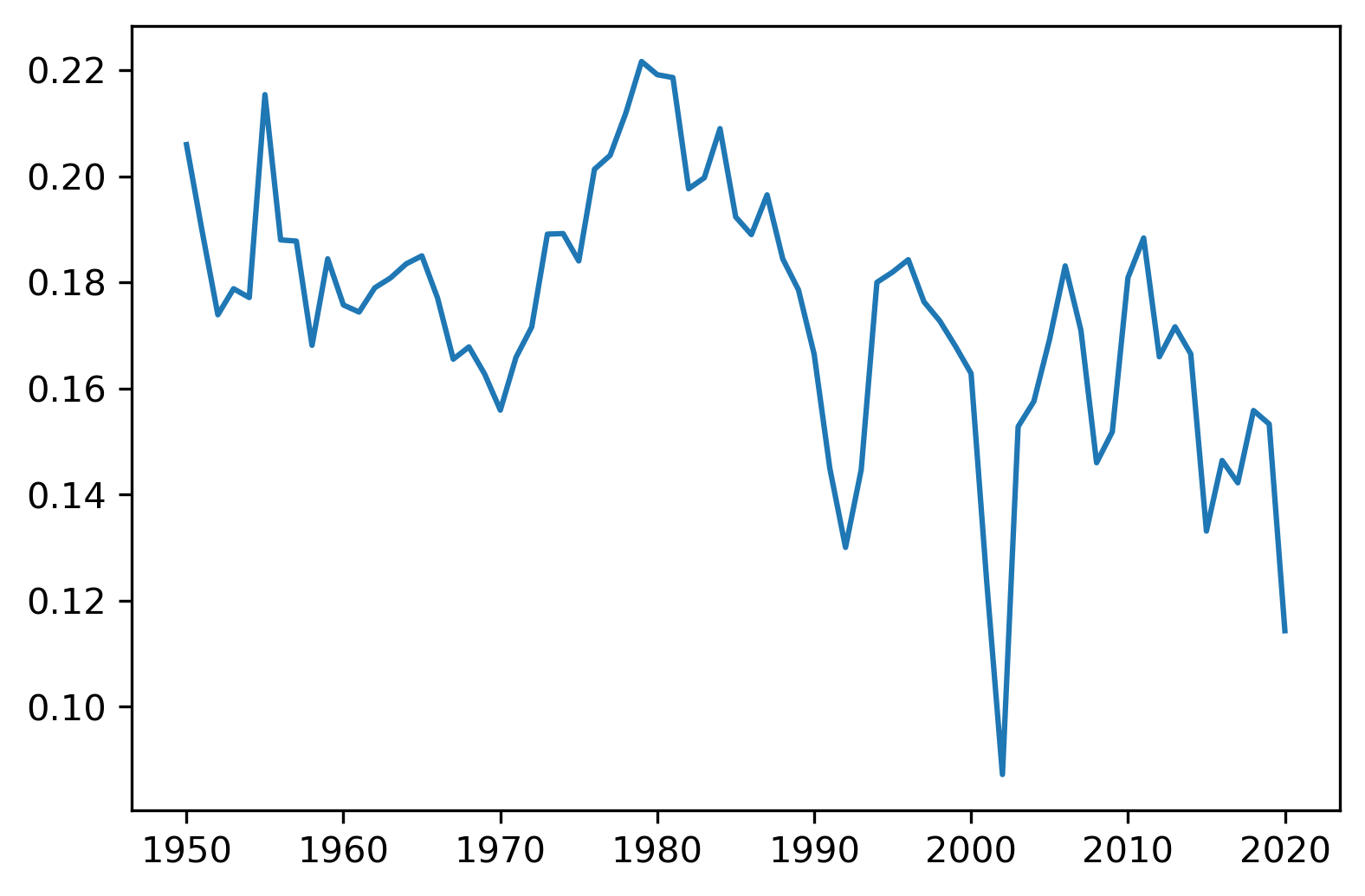}
\begin{flushleft}\singlespacing \footnotesize This plots show the ratio of aggregate Compustat profits to capital without cash. In other words, this plot shows $X / (K - \text{cash})$, where $X$ are profits, and $K$ is capital. 
\end{flushleft}
\end{figure}

\begin{figure}[!h]
\caption{Profit to Goodwill Ratio} \label{fig:profit_goodwill_ratio}
\bigskip
\centering
\includegraphics[scale=1.0]{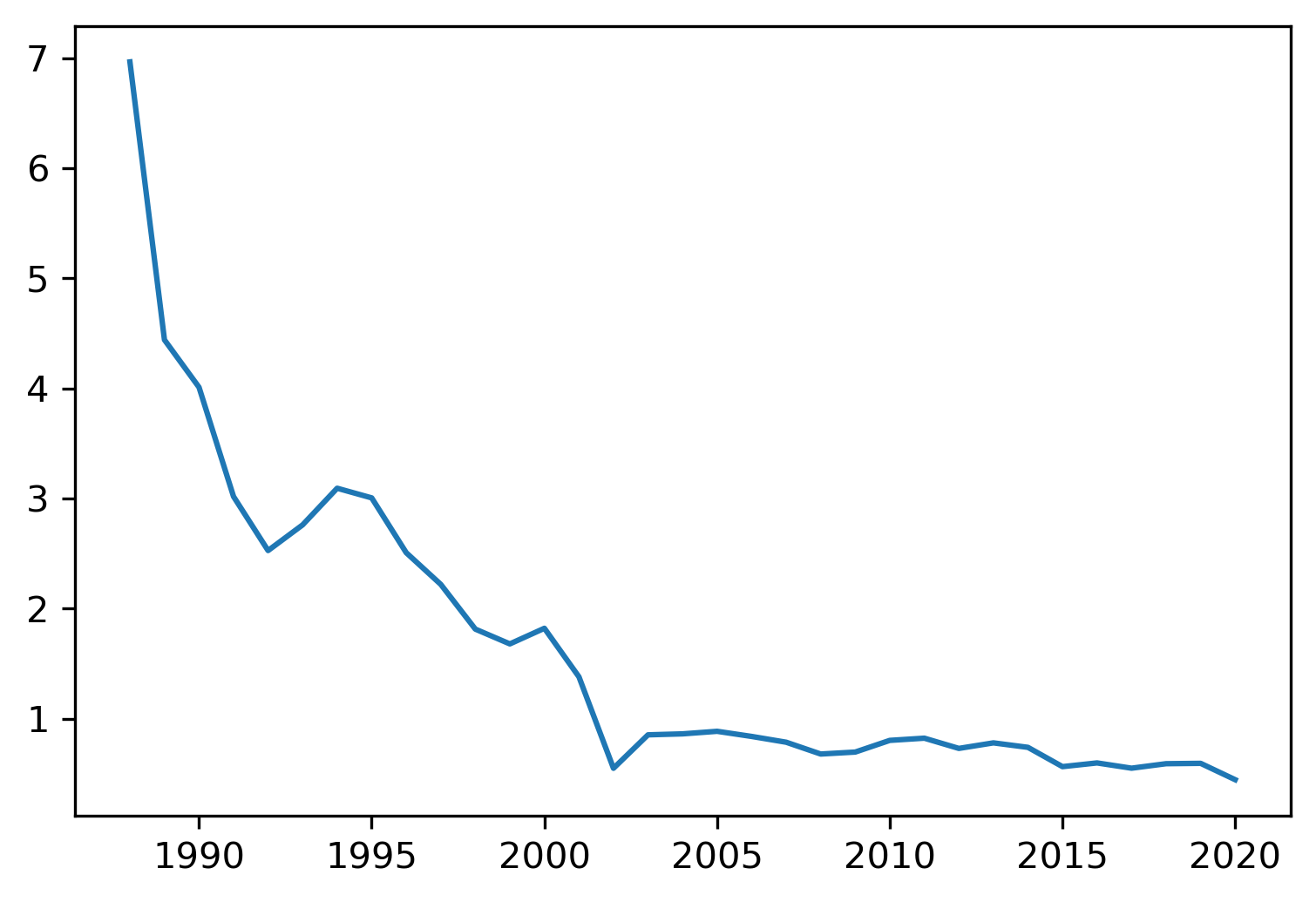}
\begin{flushleft}\singlespacing \footnotesize This plots show the ratio of aggregate Compustat profits to goodwill. 
\end{flushleft}
\end{figure}

\clearpage
\section{Tables}
\renewcommand{\thetable}{\thesection.\arabic{table}}
\setcounter{table}{0}

\begin{table}[!h]
\caption{Compustat Filters}
\label{tbl:filters}
\centering
\scalebox{1.0}{
\begin{tabular}{@{\extracolsep{5pt}}lccc}
\\[-1.8ex]\hline
\hline \\[-1.8ex]
\\[-1.8ex] Filter & Book Capital (Billions Dollars) & Fraction of Sample \\
\hline \\[-1.8ex]
currency $=$ dollars & 10,793 & 2.87\\
drop finance and utilities & 6,240 & 1.659\\
foreign\_code $=$ USA & 4,184 & 1.113\\
EXCHG $=$ 11 or EXCHG $=$ 14 & 3,965 & 1.054\\
$K$, $V$, and $X$ not missing & 3,760 & 1\\
\hline \hline \\[-1.8ex]
\end{tabular}}
\begin{flushleft}\singlespacing \footnotesize This table shows how the sample shrinks as different filters are applied to the data in order to match the US National Income and Product Accounts (NIPA) data. The left column describes the filters applied to the data, the middle column shows the average aggregate capital amount in billions of nominal dollars after the filter is applied, and right column shows the fraction of the average aggregate capital compared to the sample (the middle column divided by the last row of the middle column). The filters are described here. The currency variable is required to be US dollars (CURD $=$ USD). Finance (NAICS 520000 through 529999) and utilities (NAICS 220000 through 229999) are dropped. Only US companies are considered ($FIC = USA$, $CURCD = USD$, as well as $EXCHG = 11$ or $ECHG = 14$). Finally, only observations with non-missing capital, market capital, and profit variables are kept. 
\end{flushleft}
\end{table}

\section{Measuring the Return on Capital}
\label{app:irr}

\cite{FamaFrench1999} discuss two types of returns: (1) the internal rate of return (IRR) on cost and (2) the internal rate of return on value. The IRR on cost, $r_{C}$, is a measure of the return that comes from buying capital at cost, using it build a firm and earn its income flows, and selling the firm at market value. The IRR on value, $r_V$, is the measure of the return that comes from buying a firm at market value, earning its income flows, and selling the firm at market value.

Formally, the IRR on cost in a single period $t+1$ can be written as
\begin{equation}
    r_{C,t+1} = \frac{X_{t+1} - I_{t+1} + V_{t+1}}{K_t} - 1
\end{equation}
where $K$ is value of capital, $V$ is the market value of the firm, $X$ is aggregate cash earnings (after taxes but before interest and depreciation) and $I$ is aggregate gross investment. Likewise, the IRR on value can be written as
\begin{equation}
    r_{V,t+1} = \frac{X_{t+1} - I_{t+1} + V_{t+1}}{V_t} - 1
\end{equation}
\cite{FamaFrench1999} use the book value of equity and debt to measure capital $K$ and the market value of equity and debt to measure value $V$. These returns are measured not over a single period, but rather over many years. As the authors discuss, this alleviates measurement error as long as these returns are stationary. 

In contrast, we are interested in the time-varying dynamics of profits. Measuring these quantities over the entire sample bars us examining these time-varying dynamics. As a result, we will look at shorter time periods. However, if we measure the IRR on cost in a single period using the equation above, returns on cost will be very large since market-to-lagged-book ratios ($V_{t+1} / K_t$) are, in aggregate, usually significantly larger than unity. We therefore consider the following returns, which we label the return on capital at time $t+1$, $r_{K,t+1}$, defined as
\begin{equation}
    r_{K,t+1} = \frac{X_{t+1} - I_{t+1} + K_{t+1}}{K_t} - 1 .
\end{equation}
Since investment is defined as $I_{t+1} = K_{t+1} - K_t$, this expression simplifies to
\begin{equation}
     r_{K,t+1} = \frac{X_{t+1}}{K_t} .
\end{equation}

\end{document}